\documentstyle[11pt]{article}
\begin{document}
\begin{titlepage}
\hfill{UQMATH-93-03}
\vskip.3in
\begin{center}
{\huge Quantized Affine Lie Algebras and Diagonalization of Braid Generators}
\vskip.3in
{\Large M.D.Gould} and {\Large Y.-Z.Zhang}
\vskip.3in
{\large Department of Mathematics, University of Queensland, Brisbane,
Qld 4072, Australia}
\end{center}
\vskip.6in
\begin{center}
{\bf Abstract:}
\end{center}
Let $U_q(\hat{\cal G})$ be a quantized affine Lie algebra. It is proven
that the universal R-matrix $R$ of $U_q(\hat{\cal G})$ satisfies the
celebrated conjugation relation $R^\dagger=TR$ with $T$ the usual twist map.
As applications, braid generators are shown to be diagonalizable
on arbitrary tensor product modules of integrable irreducible highest
weight $U_q(\hat{\cal G})$-module and a
spectral decomposition formula for the braid generators is obtained
which is the generalization of
Reshetikhin's and Gould's forms to the present affine case.
Casimir invariants are constructed and their eigenvalues computed by means of
the spectral decomposition formula. As a by-product, an interesting
identity is found.

\end{titlepage}

\section{Introduction}
Braid generators for quantum (super)groups
are shown to be diagonalizable on any tensor product module of irreducible
highest weight (IHW) modules of the quantum (super)groups, in both case
of multiplicity-free \cite{Reshetikhin} and with multiplicity\cite{Gould}.
Such a diagonalized form is seen to be very useful in computing the
quantum group invariants such as link polynomials.
Here we continue the development and show some similar results
for quantized affine Lie algebras $U_q(\hat{\cal G})$.

After recalling, in section 2, some fundamentals on $U_q(\hat{\cal G})$,
we in section 3 show that the universal $R$-matrix of $U_q(\hat{\cal G})$
satisfies a conjugation relation.
In section 4 we show that braid generators are diagonzlizable on tensor
product modules of integrable IHW
$U_q(\hat{\cal G})$-module, and
obtain a spectral decomposition formula of the braid generators which
is the generalization of Reshetikhin's and Gould's forms \cite{Reshetikhin}
\cite{Gould} to our affine
case. Using this spectral decomposition formula we construct, in section 5,
a family of Casimir invariants and compute their eigenvalues on integrable
IHW modules; the eigenvalues are absolutely covergent
for $|q|\,>\,1$. Interestingly enough, we obtain an identity which bears
similarity to the power series expansion of a function in its absolutely
covergent region. Section 6 is devoted to some remarks.

\section{Fundamentals}
Let $A=(a_{ij})_{0\leq i,j\leq r}$ be a symmetrizable,
generalized Cartan matrix in the sense of Kac\cite{Kac}.
Let $\hat{\cal G}$ denote the affine algebra
associated with the corresponding symmetric Cartan
matrix $A_{\rm sym}=(a^{\rm sym}_{ij})=(\alpha_i,\alpha_j),~~i,j=0,1,...r$
\,,~$r$ is the rank of the corresponding finite-dimensional simple Lie
algebra. The
quantum algebra $U_q(\hat{\cal G})$ is defined to be a Hopf algebra with
generators: $\{e_i,~f_i,~q^{h_i}~(i=0,1,...,r),~q^d\}$ and relations
\cite{Drinfeld}\cite{Jimbo},
\begin{eqnarray}
&&q^h.q^{h'}=q^{h+h'}~~~~(h,~ h'=h_i~ (i=0,1,...,r),~d)\nonumber\\
&&q^he_iq^{-h}=q^{(h,\alpha_i)} e_i\,,~~q^hf_iq^{-h}=q^{-(h,\alpha_i)}
  f_i\nonumber\\
&&[e_i, f_j]=\delta_{ij}\frac{q^{h_i}-q^{-h_i}}{q-q^{-1}}\nonumber\\
&&\sum^{1-a_{ij}}_{k=0}(-1)^k e_i^{(1-a_{ij}-k)}e_je_i^{(k)}
   =0~~(i\neq j)\nonumber\\
&&\sum^{1-a_{ij}}_{k=0}(-1)^k f_i^{(1-a_{ij}-k)}f_jf_i^{(k)}
   =0~~(i\neq j)\label{relations1}
\end{eqnarray}
where
\begin{equation}
e_i^{(k)}=\frac{e^k_i}{[k]_q!},~~~f^{(k)}_i=\frac{f^k_i}{[k]_q!}
\,,~~~[k]_q=\frac{q^k-q^{-k}}{q-q^{-1}}\,,~~~[k]_q!=[k]_q[k-1]_q\cdots[1]_q
\end{equation}

The Cartan subalgebra (CSA) of $\hat{\cal G}$ is generated by $\{h_i,~
i=0,1,\cdots, r\,;d\}$. However, we will choose as the CSA of $U_q(\hat{\cal
G})$
\begin{equation}
{\cal H}={\cal H}_0\bigoplus{\bf C}\,c\bigoplus{\bf C}\,d
\end{equation}
where $c=h_0+h_\psi$,~ $\psi$ is the highest root of ${\cal G}$ and
${\cal H}_0$ is a CSA of ${\cal G}$.

The algebra $U_q(\hat{\cal G})$ is a Hopf algebra with coproduct, counit and
antipode similar to the case of $U_q(\cal G)$:
\begin{eqnarray}
&&\Delta(q^h)=q^h\otimes q^h\,,~~~h=h_i,~d\nonumber\\
&&\Delta(e_i)=q^{-h_i/2}\otimes e_i+e_i\otimes q^{h_i/2}\nonumber\\
&&\Delta(f_i)=
q^{-h_i/2}\otimes f_i+f_i\otimes q^{h_i/2}\nonumber\\
&&S(a)=-q^{h_\rho}aq^{-h_\rho}\,,~~~a=e_i,f_i,h_i,d\label{coproduct1}
\end{eqnarray}
where $\rho$ is the half-sum of the positive roots of $\hat{\cal G}$.
We have omitted the formula for counit since we do not need them.

Let $\Delta'$ be the opposite coproduct: $\Delta'=T\Delta$, where $T$ is
the twist map: $T(x\otimes y)=y\otimes x\,,~\forall x,y\in U_q(\hat{\cal G})$.
Then $\Delta$ and $\Delta'$ is related by the universal $R$-matrix $R$
in $U_q(\hat{\cal G})\otimes U_q(\hat{\cal G})$ satisfying, among others,
\begin{eqnarray}
&&\Delta'(x)R=R\Delta(x)\,,~~~~~\forall x\in U_q(\hat{\cal G})\nonumber\\
&&R^{-1}=(S\otimes I)R\,,~~~~~R=(S\otimes S)R\label{hopf}
\end{eqnarray}

We define a conjugate operation $\dagger$ and an anti-involution $\theta$ on
$U_q(\hat{\cal G})$ by
\begin{eqnarray}
&&d^\dagger=d\,,~~h_i^\dagger=h_i\,,~~e_i^\dagger=f_i\,,~~f_i^\dagger=e_i
 \,,~~~~i=0,1,...r \nonumber\\
&&\theta(q^h)=q^{-h}\,,~~\theta(e_i)=f_i\,,~~\theta(f_i)=
e_i\,,~~\theta(q)=q^{-1}
\end{eqnarray}
which extend uniquely to an algebra anti-automorphism and anti-involution on
all of $U_q(\hat{\cal G})$, respectively, so that $(ab)^\dagger=b^\dagger
a^\dagger\,,~~\theta(ab)=\theta(b)\theta(a)\,,~~\forall a,b\in U_q(\hat{\cal
G})$. Throughout this paper we will use the notations
\begin{eqnarray}
&&(n)_q=\frac{1-q^n}{1-q}\,,~~[n]_q=\frac{q^n-q^{-n}}{q-q^{-1}}\,,~~
  q_\alpha=q^{(\alpha,\alpha)}\nonumber\\
&&{\rm exp}_q(x)=\sum_{n\geq 0}\frac{x^n}{(n)_q!}\,,~~(n)_q!=
  (n)_q(n-1)_q\,...\,(1)_q\nonumber\\
&&({\rm ad}_qx_\alpha)x_\beta=[x_\alpha\,,\,x_\beta]_q=x_\alpha x_\beta -
 q^{(\alpha\,,\,\beta)}x_\beta x_\alpha
\end{eqnarray}

The representation theory of $U_q(\hat{\cal G})$ is analogous to
that of $\hat{\cal G}$ \cite{Rosso}\cite{ZG}. In particular,
classical and corresponding quantum representations have the same
weight spectrum. Following the usual convention, we denote the weight of
a representation by $\Lambda\equiv (\lambda,\kappa,\tau)$, where $\lambda$
$\in {\cal H}_0^*\subset{\cal H}^*$ is a weight of ${\cal G}$ and
$\kappa=\Lambda(c)\,,\,\tau=\Lambda(d)$. The non-degenerate form $(~,~)$
on ${\cal H}^*$ is defined by\cite{GO}
\begin{equation}
(\Lambda,\Lambda')=(\lambda,\lambda')+\kappa\,\tau'+\kappa'\,\tau
\end{equation}
for $\Lambda'\equiv (\lambda',\kappa',\tau')$. With these notations we have
\begin{equation}
\rho=(\rho_0,0,0)+g(0,1,0)
\end{equation}
where $\rho_0$ is the half sum of positive roots of ${\cal G}$ and
$2g=(\psi,\psi+2\rho_0)$.

We call
\begin{equation}
D_q[\Lambda]={\rm tr}(\pi_{\Lambda}(q^{2h_\rho}))
\end{equation}
the $q$-dimension of the integrable IHW representation
$\pi_\Lambda$: explicitly\cite{Kac}
\begin{eqnarray}
&&D_q[(\lambda,\kappa,\tau)]=q^{2g\tau}\,\bar{D}_q[(\lambda,\kappa,0)]
 \,, \nonumber\\
&&\bar{D}_q[(\lambda,\kappa,0)]=D^0_q[\lambda]\prod_{t=1}^\infty
\left (\frac{1-q^{-2t(\kappa+g)}}{1-q^{-2tg}}\right )^r
\prod_{\alpha\in \Phi_0}\prod_{t=1}^\infty\frac{1-q^{-2(\lambda+\rho_0,\alpha)
-2t(\kappa+g)}}{1-q^{-2(\rho_0,\alpha)-2tg}}\label{q-dimension1}
\end{eqnarray}
with
\begin{equation}
D^0_q[\lambda]=\prod_{\alpha\in\Phi^+_0}\frac{q^{(\lambda+\rho_0,\alpha)}
-q^{-(\lambda+\rho_0,\alpha)}}{q^{(\rho_0,\alpha)}-q^{-(\rho_0,\alpha)}}
\end{equation}
where $\Phi_0$ and $\Phi^+_0$ denote the set of roots and positive roots of
${\cal G}$, respectively. Note that the
$q$-dimension (\ref{q-dimension1}) is absolutely covergent for $|q|\,>\,1$.
Following the similar lines as in \cite{ZGB}, we can show
\vskip.1in
\noindent {\bf Proposition 1:} Let $V(\Lambda)$ be an IHW
$U_q(\hat{\cal G})$-module with highest weight $\Lambda\in D^+$,~$D^+$ satands
for the set of all dominant integral weights.
If the operator $\Gamma\in U_q(\hat{\cal G})\otimes {\rm End}
V(\Lambda)$ satisfies $\Delta_\Lambda(a)\Gamma=\Gamma\Delta_\Lambda(a)$,~~
$\forall a\in U_q(\hat{\cal G})$, where $\Delta_\Lambda=(I\otimes \pi_\Lambda)
\Delta$, then
\begin{equation}
C=(I\otimes {\rm tr})\{[I\otimes \pi_\Lambda(q^{2h_\rho})]\Gamma\}
\end{equation}
belongs to the center of $U_q(\hat{\cal G})$, i.e. $C$ is a Casimir invariant
of $U_q(\hat{\cal G})$.

\section{Conjugation Relation}
In this section we show that the universal R-matrix
$R$ satisfies a conjugation relation.
We illustrate the proof for the nowtwisted case,
$U_q(\hat{\cal G})=U_q({\cal G}^{(1)})$ and the same arguments are valid for
the twisted case.

To begin with, let $\hat{\cal G}={sl(2)}^{(1)}$. Fix a normal ordering in
the positive root system $\Delta_+$ of $sl(2)^{(1)}$:
\begin{equation}
\alpha,\,\alpha+\delta,\,...,\,\alpha+n\delta,\,...,\,\delta,\,2\delta,\,
...,\,m\delta,\,...\,,\,...\,,\,\beta+l\delta,\,...\,,\beta\label{order1}
\end{equation}
where $\alpha$ and $\beta$ are simple roots and $l,m,n\geq 0$;
$\delta=\alpha+\beta$ is the
minimal positive imaginary root. Khoroshkin and Tolstoy show \cite{KT}
that the universal $R$-matrix for $U_q(sl(2)^{(1)})$ may be written as
\begin{eqnarray}
R&=&\left ( \prod_{n\geq 0}{\rm exp}_{q_\alpha}((q-q^{-1})(E_{\alpha+n\delta}
  \otimes F_{\alpha+n\delta}))\right )\nonumber\\
  & &\cdot{\rm exp}\left ( \sum_{n>0}n[n]_{q_\alpha}^{-1}
  (q_\alpha-q_\alpha^{-1})(E_{n\delta}\otimes F_{n\delta})\right )\nonumber\\
& &\cdot\left (\prod_{n\geq 0}{\rm exp}_{q_\alpha}((q-q^{-1})(E_{\beta+n\delta}
  \otimes F_{\beta+n\delta}))\right )\cdot
  q^{\frac{1}{2}h_\alpha\otimes h_\alpha+c\otimes d+d\otimes c}\label{sl2R}
\end{eqnarray}
where $c=h_\alpha+h_\beta$ and the order in the product (\ref{sl2R}) concides
with the chosen normal order (\ref{order1});
Cartan-Weyl generators $E_\gamma\,,~F_\gamma=\theta(E_\gamma)\,,~~
\gamma\in \Delta_+$ of $U_q(sl(2)^{(1)})$ are defined by
\begin{eqnarray}
&&E_\alpha=e_\alpha q^{-h_\alpha/2}\,,~~~~E_\beta=e_\beta q^{-h_\beta/2}
  \,,~~~~
F_\alpha=q^{h_\alpha/2}f_\alpha\,,~~~~F_\beta=q^{h_\beta/2}f_\beta\nonumber\\
&&\tilde{E_\delta}=[(\alpha,\alpha)]_q^{-1}[E_\alpha,\,E_\beta]_q\,~~~~
E_{\alpha+n\delta}=(-1)^n\left ({\rm ad}\tilde{E_\delta}\right )^nE_\alpha
  \nonumber\\
&&E_{\beta+n\delta}=\left ({\rm ad}\tilde{E_\delta}\right )^nE_\beta\,,~~...
\,,~~\tilde{E}_{n\delta}=(\alpha,\alpha)]_q^{-1}[E_{\alpha+(n-1)\delta},\,
E_\beta]_q\label{cw1}
\end{eqnarray}
and
\begin{equation}
\tilde{E}_{n\delta}=\sum_{
\begin{array}{c}
k_1p_1+...+k_mp_m=n\\
0<k_1<...<k_m
\end{array}
}\frac{\left ( q^{(\alpha,\alpha)}-q^{-(\alpha,\alpha)}\right )^{\sum_ip_i-1}}
{p_1!\;...\;p_m!}(E_{k_1\delta})^{p_1}...(E_{k_m\delta})^{p_m} \label{ee1}
\end{equation}
By means of the following relations shown in \cite{ZG}
\begin{eqnarray}
&&S(E^\dagger_{\alpha+n\delta})=-q^{n(\alpha,\beta)}F_{\alpha+n\delta}\,,~~~
  S(E^\dagger_{\beta+n\delta})=-q^{n(\alpha,\beta)}F_{\beta+n\delta}\nonumber\\
&&S(F^\dagger_{\alpha+n\delta})=-q^{-n(\alpha,\beta)}E_{\alpha+n\delta}\,,~~~
  S(F^\dagger_{\beta+n\delta})=-q^{-n(\alpha,\beta)}
  E_{\beta+n\delta}\nonumber\\
&&S(\tilde{E}^\dagger_{n\delta})=-q^{n(\alpha,\beta)}\tilde{F}_{n\delta}\,,~~~
  S(E^\dagger_{n\delta})=-q^{n(\alpha,\beta)}F_{n\delta}\nonumber\\
&&S(\tilde{F}^\dagger_{n\delta})=-q^{-n(\alpha,\beta)}\tilde{E}_{n\delta}\,,~~~
  S(F^\dagger_{n\delta})=-q^{-n(\alpha,\beta)}E_{n\delta}\label{s1}
\end{eqnarray}
We now prove the following
\vskip.1in
\noindent {\bf Proposition 2:} The universal $R$-matrix for $U_q(sl(2)^{(1)})$
satisfies the conjugation relation: $R^\dagger=R^{T}$
\vskip.1in
\noindent {\bf Proof:} The univarsal $R$-matrix (\ref{sl2R}) can be written as
\begin{eqnarray}
R&=&\sum_{{\bf l},{\bf n},{\bf k}}\;
  A_{{\bf l},{\bf n},{\bf k}}(q)
  (E_\alpha)^{l_0}\,...\,
  (E_{\alpha+N\delta})^{l_N}\,...\,
  (E_\delta)^{n_1}\,...\,(E_{L\delta})^{n_L}\,...\,...\,\nonumber\\
& &\cdot  (E_{\beta+M\delta})^{k_M}\,...\,(E_\beta)^{k_0}
  \otimes (F_\alpha)^{l_0}\,...\,
  (F_{\alpha+N\delta})^{l_N}\,...\,(F_\delta)^{n_1}\,...\,(F_{L\delta})^{n_L}
  \,...\,\nonumber\\
& &\cdot \,...\, (F_{\beta+M\delta})^{k_M}\,...\,(F_\beta)^{k_0}\;
  q^{\frac{1}{2}h_\alpha\otimes h_\alpha+c\otimes d+d\otimes c}\label{1}
\end{eqnarray}
where $\{{\bf l}\}=\{l_0,l_1,...,l_N,...\},~~\{{\bf n}\}=\{n_1,n_2,...,n_L,...
\}$\,,~~$\{{\bf k}\}=\{k_0,k_1,...,k_M,...\}$;~
the constants $A_{{\bf l},{\bf n},{\bf k}}(q)$ are given by
\begin{eqnarray}
A_{{\bf l},{\bf n},{\bf k}}(q)&=&\frac{(q-q^{-1})^{l_0+l_1+...+l_N+...}}
{(l_0)_{q_\alpha}!\,...\,(l_N)_{q_\alpha}!\,...}
\frac{(q-q^{-1})^{k_0+k_1+...+k_M+...}}
{(k_0)_{q_\alpha}!\,...\,(k_M)_{q_\alpha}!\,...}\nonumber\\
& &\cdot \frac{1^{n_1}\,...\,L^{n_L}\,...\;(q-q^{-1})^{n_1+n_2+...+n_L+...}}
{[1]_{q_\alpha}^{n_1}\,...\,[L]_{q_\alpha}^{n_L}\,...\;n_1!\,...\,n_L!\,...}
\end{eqnarray}
{}From (\ref{1}) we deduce
\begin{eqnarray}
(S\otimes S)R^\dagger&=&\sum_{{\bf l},{\bf n},{\bf k}}\;
  {A}_{{\bf l},{\bf n},{\bf k}}(q)
 S(E^\dagger_\alpha)^{l_0}\,...\,
  S(E^\dagger_{\alpha+N\delta})^{l_N}\,...\,
  S(E^\dagger_\delta)^{n_1}\,...\,S(E^\dagger_{L\delta})^{n_L}\,...\,...\,
  \nonumber\\
& &\cdot  S(E^\dagger_{\beta+M\delta})^{k_M}\,...\,S(E^\dagger_\beta)^{k_0}
  \otimes S(F^\dagger_\alpha)^{l_0}\,...\,
  S(F^\dagger_{\alpha+N\delta})^{l_N}\,...\,S(F^\dagger_\delta)^{n_1}\,...\,
  S(F^\dagger_{L\delta})^{n_L}\nonumber\\
& &\cdot\,...\,...\,  S(F^\dagger_{\beta+M\delta})^{k_M}\,...\,
  S(F^\dagger_\beta)^{k_0}\;
  q^{\frac{1}{2}h_\alpha\otimes h_\alpha+c\otimes d+d\otimes c}
\end{eqnarray}
which gives, with the help of (\ref{s1})
\begin{equation}
(S\otimes S)R^\dagger=R^{T}
\end{equation}
Hence by (\ref{hopf})
\begin{equation}
R^\dagger=(S^{-1}\otimes S^{-1})R^{T}=R^{T},
\end{equation}
as required. ~~~~$\Box$

Next we come to general case: $\hat{\cal G}={\cal G}^{(1)}$.
Fix some order in the positive
root system $\Delta_+$ of ${\cal G}^{(1)}$, which satisfies
an additional condition,
\begin{equation}
\alpha+n\delta~\leq~k\delta~\leq~(\delta-\beta)+l\delta\label{order2}
\end{equation}
where $\alpha\,,~\beta\,\in~\Delta^0_+$\,,~~$\Delta_+^0$ is the positive
root system of ${\cal G}$\,;~$k\,,\,l\,,\,n\,\geq\,0$ and
$\delta$ is the minimal positive imaginary root. Then
the universal $R$-matrix $U_q({\cal G}^{(1)})$
may be written in the following form \cite{KT},
\begin{eqnarray}
R&=&\left (\prod_{\gamma\in \Delta_+^{\rm re}\,,\,\gamma<\delta}~~{\rm exp}_{q_
\gamma}\left (\frac{q-q^{-1}}{C_\gamma(q)}E_\gamma\otimes F_\gamma\right )
\right )\nonumber\\
& &\cdot {\rm exp}\left (\sum_{n>0}\sum^r_{i,j=1}
C^n_{ij}(q)(q-q^{-1})(E^{(i)}_{n\delta}\otimes F^{(j)}_{n\delta})
\right )\nonumber\\
& &\cdot \left (\prod_{\gamma\in \Delta_+^{\rm re}\,,\,\gamma>\delta}~~
{\rm exp}_{q_
\gamma}\left (\frac{q-q^{-1}}{C_\gamma(q)}E_\gamma\otimes F_\gamma\right )
\right )\cdot q^{\sum^r_{i,j=1}\,(a^{-1}_{\rm sym})^{ij}h_i\otimes h_j
+c\otimes d+d\otimes c}\label{generalR}
\end{eqnarray}
where $c=h_0+h_{\psi}$,~$\psi$ is the highest root of ${\cal G}$ and
Cartan-Weyl generators $E_\gamma$ and $F_\gamma=\theta(E_\gamma)\,,~~\gamma
\in \Delta_+$  are defined similarly as to (\ref{cw1} , \ref{ee1}) \cite{KT};
$(C^n_{ij}(q))\,,~~i,j=1,2,...,r$, is the inverse of the matrix
$(B^n_{ij}(q))\,,~~i,j=1,2,...,r$ with
\begin{equation}
B^n_{ij}(q)=(-1)^{n(1-\delta_{ij})}n^{-1}\frac{q^n_{ij}-q^{-n}_{ij}}
{q_{j}-q^{-1}_{j}}\frac{q-q^{-1}}{q_{i}-q^{-1}_{i}}
\,,~~~~q_{ij}=q^{(\alpha_i,\alpha_j)}\,,~~~q_i\equiv q_{\alpha_i}
\end{equation}
and $C_\gamma(q)$ is a normalizing constant defined by
\begin{equation}
[E_\gamma\,,\,F_\gamma]=\frac{C_\gamma(q)}{q-q^{-1}}\left ( q^{h_\gamma}
-q^{-h_\gamma}\right )\,,~~~~\gamma\in\Delta^{\rm re}_+
\end{equation}
The order in the product of the $R$-matrix concides with the chosen normal
ordering (\ref{order2}) in $\Delta_+$.
One can show \cite{ZG} that for any $\alpha\in \Delta^0_+$,
\begin{eqnarray}
&&S(E^\dagger_{\alpha+n\delta})=-q^{(\alpha,\alpha-2\rho)/2-n(\delta,\rho)}
F_{\alpha+n\delta}\nonumber\\
&&S(F^\dagger_{\alpha+n\delta})=-q^{-(\alpha,\alpha-2\rho)/2+n(\delta,\rho)}
E_{\alpha+n\delta}\nonumber\\
&&S(E^\dagger_{\delta-\alpha+n\delta})=-q^{(\delta-\alpha,\delta-\alpha-2\rho)
/2-n(\delta,\rho)}
F_{\delta-\alpha+n\delta}\nonumber\\
&&S(F^\dagger_{\delta-\alpha+n\delta})=-q^{-(\delta-\alpha,\delta-\alpha-2\rho)
/2+n(\delta,\rho)}
E_{\delta-\alpha+n\delta}\nonumber\\
&&S(\tilde{E}^{(i)\dagger}_{n\delta})=-q^{-n(\delta,\rho)}
\tilde{F}^{(i)}_{n\delta}\,.~~~~~
S(\tilde{F}^{(i)\dagger}_{n\delta})=-q^{n(\delta,\rho)}
\tilde{E}^{(i)}_{n\delta}\nonumber\\
&&S(E^{(i)\dagger}_{n\delta})=-q^{-n(\delta,\rho)}F^{(i)}_{n\delta}\,,~~~~~
S(F^{(i)\dagger}_{n\delta})=-q^{n(\delta,\rho)}E^{(i)}_{n\delta}\label{s2}
\end{eqnarray}
We are now in a position to state
\vskip.1in
\noindent {\bf Proposition 3:}
The universal $R$-matrix for $U_q({\cal G}^{(1)})$
satisfies the conjugation relation: $R^\dagger=R^{T}$
\vskip.1in
\noindent {\bf Proof:} This is proven exactly as for proposition 2 by means of
(\ref{s2}).~~~$\Box$
\vskip.1in
\noindent{\bf Remark 1:} Proposition 2 and 3 are also valid for {\em twisted}
quantized affine Lie algebras.

\section{Evaluation of Braid Generators}
It is a well established fact that for
quasitriangular Hopf algebras, there exists a distinguished element
\cite{Drinfeld}\cite{FR}
\begin{equation}
u=\sum_tS(b_t)a_t\label{u}
\end{equation}
where $a_t$ and $b_t$ are coordinates of the universal $R$-matrix
\begin{equation}
R=\sum_ta_t\otimes b_t\label{r}
\end{equation}
One can show that $u$ has inverse
\begin{equation}
u^{-1}=\sum_tS^{-2}(b_t)a_t\label{u-1}
\end{equation}
and satisfies
\begin{eqnarray}
&&S^2(a)=uau^{-1}\,,~~~\forall a\in U_q(\hat{\cal G})\nonumber\\
&&\Delta(u)=(u\otimes u)(R^TR)^{-1}
\end{eqnarray}
where $R^T=T(R)$.  It is easy to check that
$v=uq^{-2h_\rho}$ belongs to the center
of $U_q(\hat{\cal G})$ and satisfies
\begin{equation}
\Delta(v)=(v\otimes v) (R^TR)^{-1}\label{vv1}
\end{equation}
Moreover, on an irreducible representation with highest weight
$\Lambda\equiv (\lambda,\kappa,\tau)\in D^+$, the Casimir operator $v$
takes the eigenvalue \cite{ZG}
\begin{equation}
\chi_\Lambda=q^{-(\Lambda,\Lambda+2\rho)}\label{chi1}
\end{equation}

Let $V\equiv V(\lambda,\kappa,0)$ (where without loss generality we have
set $\tau=0$) and
$P$ be the permutation operator on $V\otimes V$ defined by
$P(|\mu>\otimes |\nu>)=|\nu>\otimes |\mu>\,,~\forall |\mu>\,,\,|\nu>\in
V$ and Let
\begin{equation}
\sigma=PR\,~~~~~\in\,{\rm End}(V\otimes V)\label{pr}
\end{equation}
Here and in what follows we regard elements of $U_q(\hat{\cal G})$ as
operators on $V$. Then (\ref{hopf}) is equivalent to
\begin{equation}
\sigma\Delta(a)=\Delta(a)\sigma\,~~~~\forall a\in U_q(\hat{\cal G})\label{11}
\end{equation}
which means that $\sigma$ is an operator obeying the condition of proposition
1. Furthermore,
\vskip.1in
\noindent{\bf Proposition 4:} $\sigma$ is self-adjoint and thus may be
diagonalized.
\vskip.1in
\noindent {\bf Proof:} It follows from propositions 2, 3 and remark 1:
\begin{equation}
\sigma^\dagger=(PR)^\dagger=R^\dagger P=R^TP=P\cdot PR^TP=PR=\sigma~~~~~\Box
\end{equation}
\vskip.1in
Recall that ${\rm lim}_{q\rightarrow 1}\;\sigma=P$ and $P$ is diagonalizable
on $V\otimes V$ with eigenvalues $\pm 1$. Following \cite{Gould}, we define
the subspaces
\begin{equation}
W_\pm=\{w\in V\otimes V\,|\, {\rm lim}_{q\rightarrow 1}\sigma w=\pm w\}
\end{equation}
Since $\sigma$ is self-adjoint we may clearly write
\begin{equation}
V\otimes V=W_+\bigoplus W_-
\end{equation}
Let $P[\pm]$ denote the projection operators defined by
\begin{equation}
P[\pm](V(\lambda,\kappa,0)\otimes V(\lambda,\kappa,0))=W_\pm
\end{equation}
Since $\sigma$ is an $U_q(\hat{\cal G})$-invariant each subspace $W_\pm$
determines a $U_q(\hat{\cal G})$-module and $P[\pm]$ commute with the action
of $U_q(\hat{\cal G})$. On the other hand, it is shown \cite{ZG} that the
tensor product $V\otimes V$ of integrable IHW $U_q(\hat{\cal G})$-module
$V(\lambda,\kappa,0)$ is completely reducible and the irreducible components
are integrable highest weight representations. This means that
we may decompose the tensor product $V\otimes V$ according to
\begin{equation}
V(\lambda,\kappa,0)\otimes V(\lambda,\kappa,0)=\bigoplus_{(\mu,2\kappa,-s)\in
D^+}\bigoplus_{s\geq 0}\,\bar{V}(\mu,2\kappa,-s)\label{decom1}
\end{equation}
where the sum on $\mu$ is finite and
\begin{equation}
\bar{V}(\mu,2\kappa,-s)=V(\mu,2\kappa,-s)\bigoplus\,\cdots\,\bigoplus
V(\mu,2\kappa,-s)~~~~~~~(m_{\mu,s}~ {\rm terms})
\end{equation}
with $m_{\mu,s}$ being the multiplicity of the module $V(\mu,2\kappa,-s)$ in
the tensor product decomposition (\ref{decom1}). Let $P[\mu,s]$ be the central
projections:
\begin{equation}
P[\mu,s] (V(\lambda,\kappa,0)\otimes V(\lambda,\kappa,0))=\bar{V}
(\mu,2\kappa,-s)
\end{equation}
We then deduce the following decompositions:
\begin{equation}
W_\pm=\bigoplus_{(\mu,2\kappa,-s)\in D^+}\bigoplus_{s\geq 0}\,\bar{V}_\pm
(\mu,2\kappa,-s)
\end{equation}
where
\begin{equation}
\bar{V}_\pm (\mu,2\kappa,-s)=P[\pm]\bar{V}(\mu,2\kappa,-s)
\end{equation}
Let
\begin{equation}
P[\mu,s;\pm]=P[\mu,s]P[\pm]=P[\pm]P[\mu,s]
\end{equation}
then
\begin{equation}
\bar{V}_\pm (\mu,2\kappa,-s)=P[\mu,s;\pm] (V(\lambda,\kappa,0)\otimes
V(\lambda,\kappa,0))=P[\mu,s]W_\pm
\end{equation}

We now observe
\begin{equation}
\sigma^2=PRP\cdot R=R^TR=(v\otimes v)\Delta(v^{-1})
\end{equation}
where we have used (\ref{vv1}) in the last step. It then follows from
(\ref{chi1}) that on the submodule $\bar{V}_\pm (\mu,2\kappa,-s)$ the
$\sigma^2$ has the eigenvalue
\begin{equation}
\chi_{(\mu,2\kappa,-s)}(\sigma^2)=\chi_{(\lambda,\kappa,0)}(v)
\chi_{(\lambda,\kappa,0)}(v)\chi_{(\mu,2\kappa,-s)}(v^{-1})=
q^{(\mu,\mu+2\rho_0)-2(\lambda,\lambda+2\rho_0)-2s(2\kappa+g)}
\end{equation}
Since $\sigma$ is self-adjoint the above equation implies
\begin{equation}
\chi_{(\mu,2\kappa,-s)}(\sigma)=\pm
q^{(\mu,\mu+2\rho_0)/2-(\lambda,\lambda+2\rho_0)-s(2\kappa+g)}
\end{equation}
on $\bar{V}_\pm (\mu,2\kappa,-s)$, respectively. We thus arrive at the
following spectral decomposition formula for $\sigma$ and its powers:
\begin{eqnarray}
\sigma^l&=&q^{-l(\lambda,\lambda+2\rho_0)}\sum_{(\mu,2\kappa,-s)\in D^+}
\sum_{s=0}^\infty q^{l(\mu,\mu+2\rho_0)/2-sl(2\kappa+g)}\nonumber\\
 & &\cdot \left ( P[+]+(-1)^{l}P[-]\right ) P[\mu,s],~~~~
 l\in {\bf Z}\label{sigma1}
\end{eqnarray}
This representation of the braid generators are the generalization
of the ones in \cite{Reshetikhin}\cite{Gould} to the case at hand.

It should be pointed that that although $\sigma$ and its
eigenvalue are bounded above for $|q|\,>\,1$, $\sigma^{-1}$ and its
eigenvalue are not.

\section{Casimir Operators}
It follows from (\ref{11}) and proposition 1 that operators
\begin{equation}
C^{(l)}=(I\otimes{\rm tr})\{[I\otimes\pi_\Lambda(q^{2h_\rho})]\sigma^l\}
\,,~~~~~l\in {\bf Z}^+
\end{equation}
are Casimir invariants acting on $V(\lambda)$. Their eigenvalues, denoted
$c_\Lambda^{(l)}$, are given by
\vskip.1in
\noindent{\bf Proposition 5:}
\begin{equation}
c_\Lambda^{(l)}=q^{-l(\lambda,\lambda+2\rho_0)}\sum_{(\mu,2\kappa,-s)\in D^+}
\sum_{s=0}^\infty (m^+_{\mu,s}+(-1)^lm^-_{\mu,s})q^{l(\mu,\mu+2\rho_0)/2-
sl(2\kappa+g)}\frac{D_q[(\mu,2\kappa,-s)]}{D_q[(\lambda,\kappa,0)]}
\,,~~~~l\in {\bf Z}^+\label{eigenvalue1}
\end{equation}
where $m^\pm_{\mu,s}\in
{\bf Z}^+$ are the mutliplicity of $V(\mu, 2\kappa,-s)$ in $W_\pm$,
respectively, so that $m_{\mu,s}=m^+_{\mu,s}+m^-_{\mu,s}$. The r.h.s. of
(\ref{eigenvalue1}) is absolutely covergent for $|q|\,>\,1$.
\vskip.1in
\noindent{\bf Proof:} This can be proven as in \cite{Gould} with the help of
the decomposition (\ref{sigma1}).~~~$\Box$
\vskip.1in
On the other hand, for $l=1$ we can prove by direct computation
\begin{equation}
C^{(1)}=(I\otimes {\rm tr})\{[I\otimes\pi_\Lambda(q^{2h_\rho})]
  \sigma\}=c^{(1)}_\Lambda I\label{other c}
\end{equation}
with
\begin{equation}
c^{(1)}_\Lambda=q^{(\Lambda,\Lambda+2\rho)}
\end{equation}
This is done as follows. Let $|\Lambda>$ denote the highest weight vector
of $V(\Lambda)$ with the highest weight $\Lambda$. Consider
\begin{equation}
<\Lambda|q^{2h_\rho}C^{(1)}|\Lambda>=q^{2(\Lambda,\rho)}{c}^{(1)}_\Lambda
\end{equation}
then by (\ref{pr}) and (\ref{r})
\begin{eqnarray}
{\rm the~ l.h.s.}&=&\sum_\nu <\Lambda\otimes\nu|\Delta(q^{2h_\rho})P
\sum_t(a_t\otimes b_t)|\Lambda\otimes\nu>=\sum_{\nu,t}<\nu\otimes\lambda
|q^{2h_\rho}a_t\otimes q^{2h_\rho}b_t|\Lambda\otimes\nu>\nonumber\\
&=&\sum_t <\Lambda|q^{2h_\rho}b_tq^{2h_\rho}a_t
|\Lambda>=\sum_t<\Lambda|q^{4h_\rho}q^{-2h_\rho}b_tq^{2h_\rho}
a_t|\Lambda>\nonumber\\
&=&q^{4(\Lambda,\rho)}<\Lambda|S^{-2}(b_t)a_t)|\Lambda>
=q^{4(\Lambda,\rho)}<\Lambda|u^{-1}|\Lambda>
=q^{2(\Lambda,\rho)}<\Lambda|
u^{-1}q^{-2h_\rho}|\Lambda>\nonumber\\
&=&q^{2(\Lambda,\rho)}<\Lambda|v^{-1}|\Lambda>
=q^{2(\Lambda,\rho)+(\Lambda,\Lambda+2\rho)}
\end{eqnarray}
where use has been made of (\ref{chi1}) and
\begin{equation}
q^{-2h_\rho}b_tq^{2h_\rho}=S^{-2}(b_t)
\end{equation}

By comparing (\ref{other c}) with (\ref{eigenvalue1}), we obtain the
interesting identity
\begin{equation}
q^{2(\lambda,\lambda+2\rho_0)}=\sum_{(\mu,2\kappa,-s)\in D^+}
\sum_{s=0}^\infty (m^+_{\mu,s}-m^-_{\mu,s})q^{(\mu,\mu+2\rho_0)/2-
s(2\kappa+g)}\frac{D_q[(\mu,2\kappa,-s)]}{D_q[(\lambda,\kappa,0)]}\label{id}
\end{equation}
Note that the r.h.s of the above relation is absolutely covergent for
$|q|\,>\,1$. Eq.(\ref{id})  bears similarity to the power series expansion
of a function in its absolutely covergent region.

\section{Remarks}
In ref.\cite{Reshetikhin}\cite{ZGB} the spectral decomposition formulae
similar to (\ref{sigma1}) are applied to define link polynomials explicitly
for a general irreducible representation of a quantum (super)group.
However in the case at hand care must be taken in order to avoid
divergence issues since the inverse of the braid generator is unbounded
above. It would be of interest to apply the results above to determine
generalized link polynomials associated with the integrable IHW
representations of the present affine case. Work towards this goal is
under investigation.
\vskip.3in
\begin{center}
{\bf Acknowledgements:}
\end{center}
The financial support from the
Australian Research Council is gratefully acknowledged.

\end{document}